\documentclass{article}
\usepackage{spconf,amsmath,graphicx}

\usepackage[Export]{adjustbox}
\usepackage{multirow}
\usepackage{array}
\usepackage[colorlinks=true, urlcolor=blue, linkcolor=red]{hyperref}
\usepackage{float}
\usepackage{tabularx}

\usepackage{listings}
\usepackage{fancyvrb}
\usepackage{framed}
\newlength\rowheight
\newlength\colwidth
\newlength\blockspace
\newlength\smallspace
\newlength\multirowcor

\title{Integrating Text-to-Music Models with Language Models:\\
Composing Long Structured Music Pieces}
\name{Lilac Atassi}
\address{University of California, San Diego}
\begin{document}
%
\maketitle
\begin{abstract}
Recent music generation methods based on transformers have a context window of up to a minute. The music generated by these methods is largely unstructured beyond the context window. With a longer context window, learning long-scale structures from musical data is a prohibitively challenging problem. This paper proposes integrating a text-to-music model with a large language model to generate music with form. The papers discusses the solutions to the challenges of such integration. The experimental results show that the proposed method can generate 2.5-minute-long music that is highly structured, strongly organized, and cohesive.
\end{abstract}
\section{Introduction}
\label{sec:intro}

The new wave of generative models has been explored in the literature to generate music. Jukebox \cite{dhariwal2020jukebox} is based on Hierarchical VQ-VAEs \cite{razavi2019generating} to generate multiple minutes of music. Jukebox is one of the earliest purely learning-based models that could generate longer than one minute of music with some degree of structural coherence. Notably, the authors mention that the generated music at a small scale of multiple seconds is coherent, and at a larger scale, beyond one minute, it lacks musical form.

\begin{figure}[!t]
\begin{center}
\includegraphics[width=0.99\columnwidth]{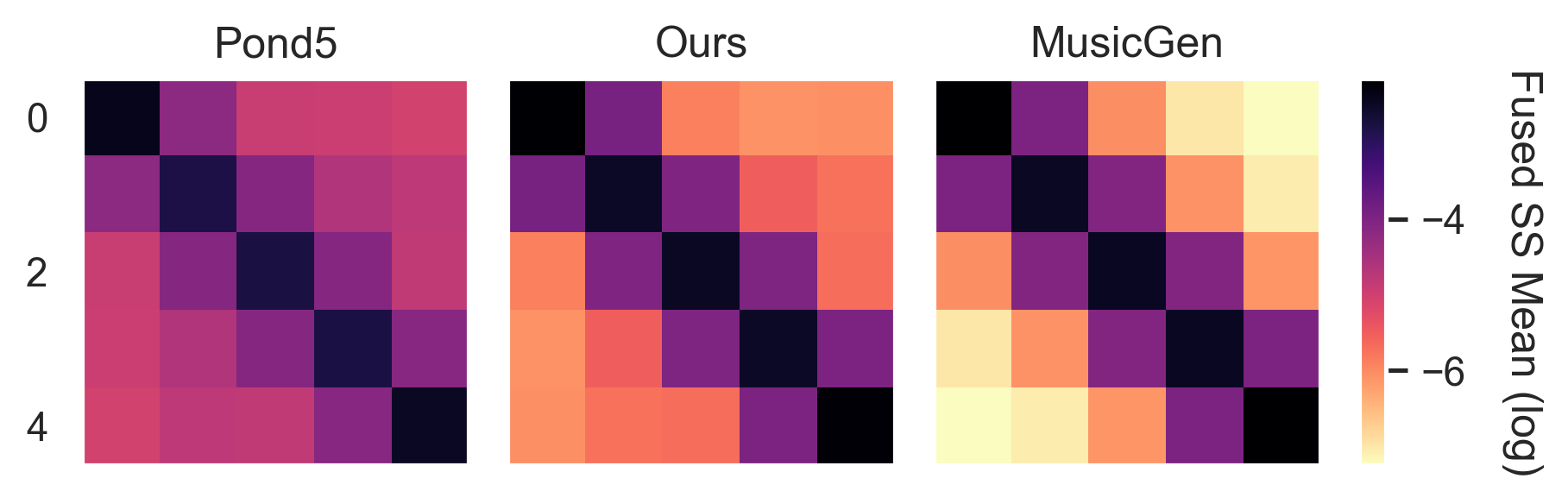}
\includegraphics[width=0.99\columnwidth]{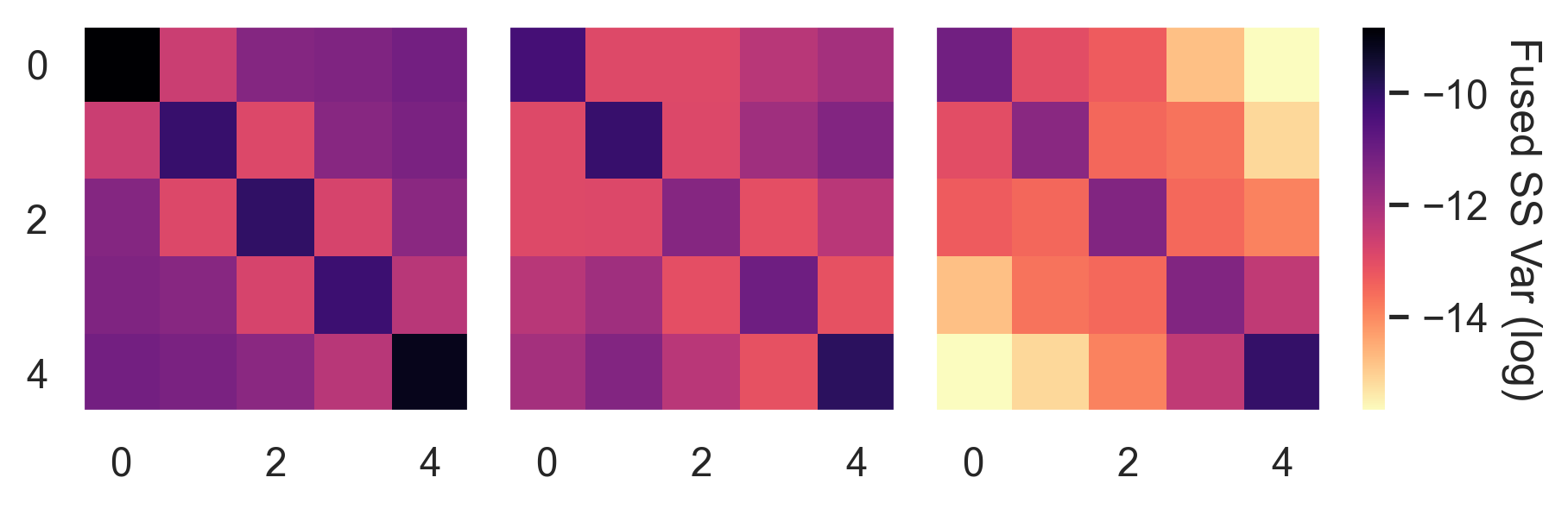}

\caption{The mean (top row) and the variance (bottom row) of the fused self-similarity (SS) matrices estimated by 100 samples from Pond5, generated by our method, and by MusicGen. The SS matrices are downsampled to $5\times5$. The results indicate that, compared to MusicGen, our method produces samples that more closely resemble the Pond5 samples in terms of long-term temporal consistency and the diversity of recurring sections.}
\label{fig:mus_objective_eval}
\end{center}
\end{figure}

Music Transformer \cite{huang2018music} adapted the transformer architecture for music generation, capable of processing the number of input tokens equivalent to up to one minute of music. More recent works on transformers and autoregressive models first use an efficient encoder to compress the audio. In the compressed space, the transformer model is trained to generate new tokens. Then, the generated tokens are converted into audio by a decoder. MusicGen \cite{copet2024simple} follows this approach.

The generative models in the literature have been applied to both symbolic \cite{wang2024diffuseroll, mittal2021symbolic, lupker2021score} and audio music \cite{baoueb2024specdiff,hawthorne2022multi}. Most of the recent larger models are trained on audio \cite{copet2024simple,agostinelli2023musiclm}, as collecting a large training set of music audio is more readily available than symbolic music. 

Earlier works on generative music models, such as JukeBox, attempted to optimize the model to process long sequences of multiple minutes in the hope that the model would learn musical structures and forms at all scales. However, none of the models in the literature has demonstrated musical structure at a large scale and lack even simple musical forms. Section 2 presents the arguement that it is impractical to have a sufficiently large training dataset. Using Large Language Models (LLMs), this papers proposes an approach to generate music with organized structures on larger scales. Section 3 presents a brief review of MusicGen. Sections 4 discusses the proposed approach to integrate MusicGen with ChatGPT~\cite{achiam2023gpt}. The experiments and evaluation of our method are presented in Section 5. Section 6 concludes the paper.

\section{Learning Musical Form}

\begin{figure*}[htb!]
  \setlength\rowheight{2.6cm}
  \setlength\colwidth{.18\textwidth}
  \setlength\blockspace{0cm}
  \setlength\smallspace{0.1cm}
  \setlength\multirowcor{-5mm}
  \adjustboxset{width=\colwidth,height=\rowheight,valign=c}
  \centering
  \renewcommand{\arraystretch}{.5} 
  \setlength{\tabcolsep}{0mm}  
  
\begin{tabularx}{\textwidth}{LFFFFFF}
\hspace{.35cm}
\rotatebox[origin=t]{90}{Dall-E}
& \includegraphics[width=0.15\textwidth]{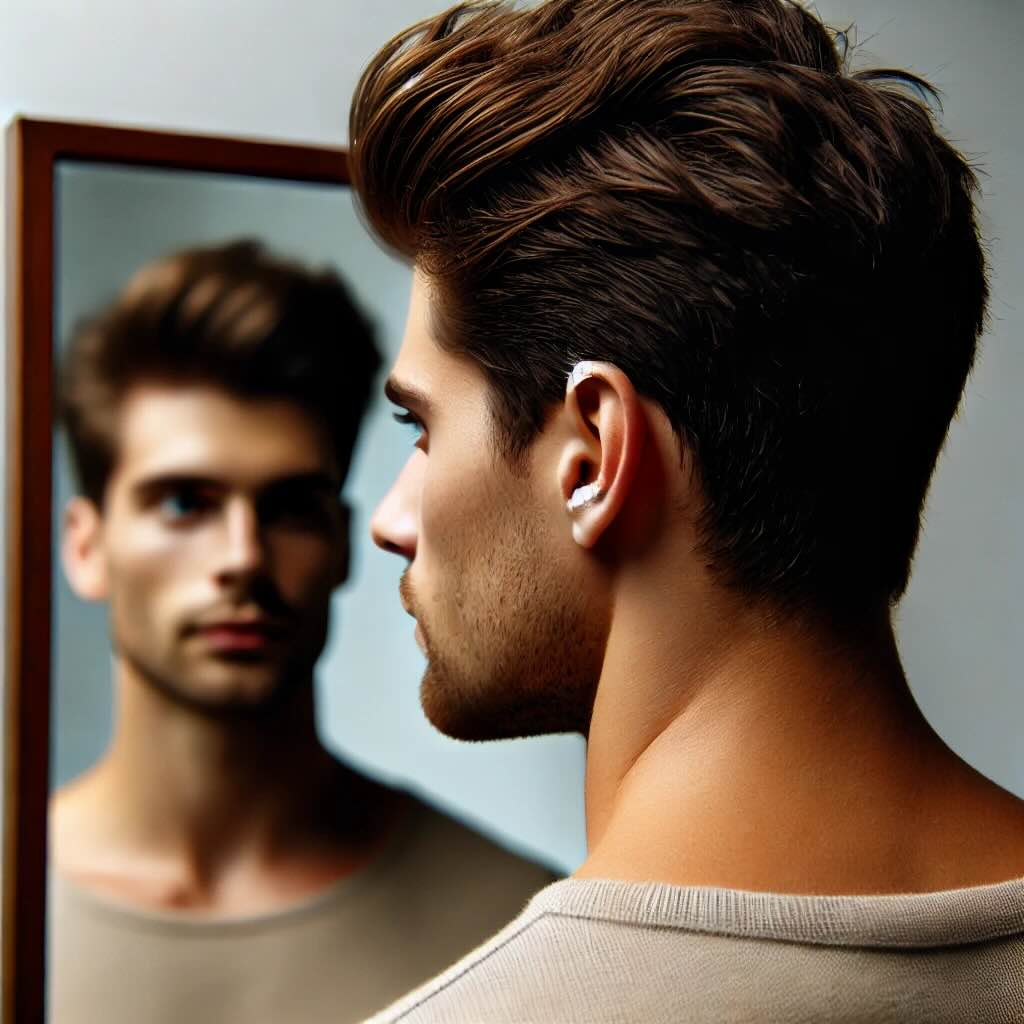}
& \includegraphics[width=0.15\textwidth]{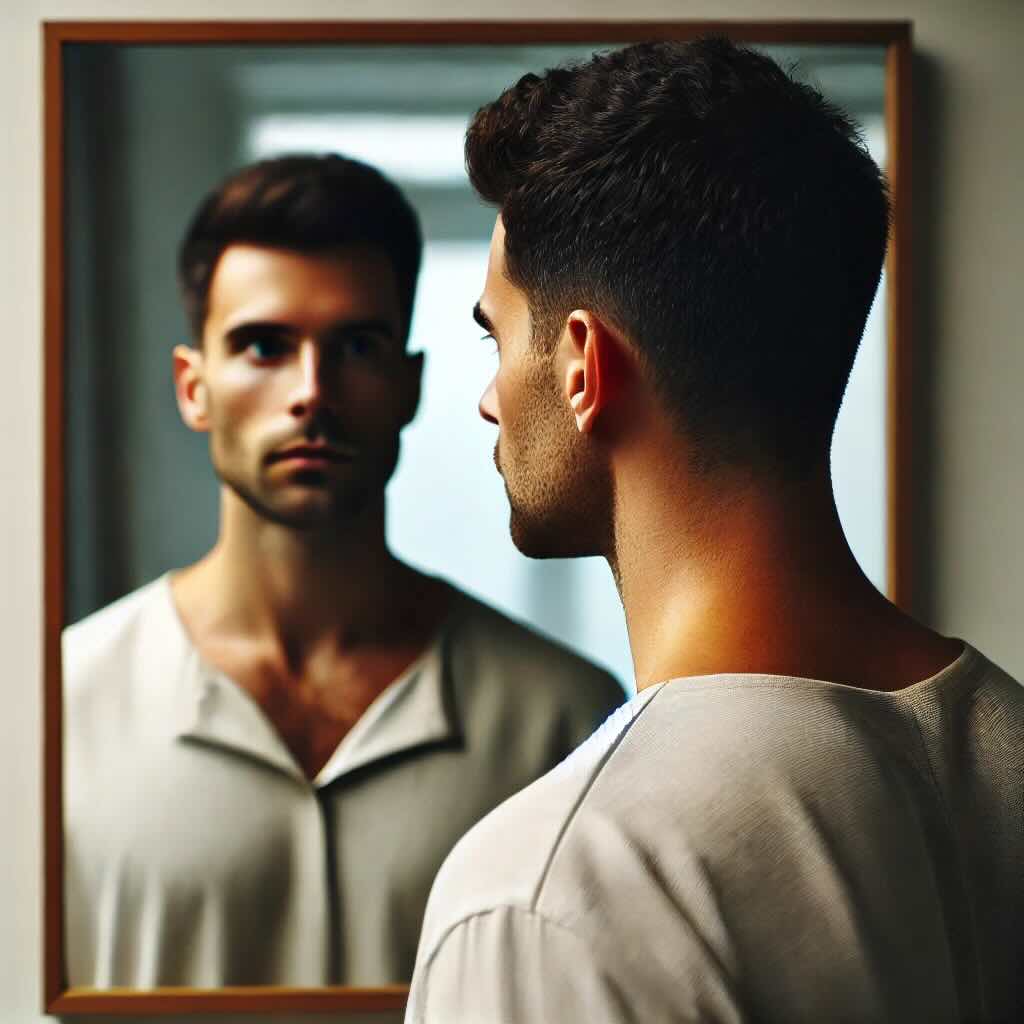}
& \includegraphics[width=0.15\textwidth]{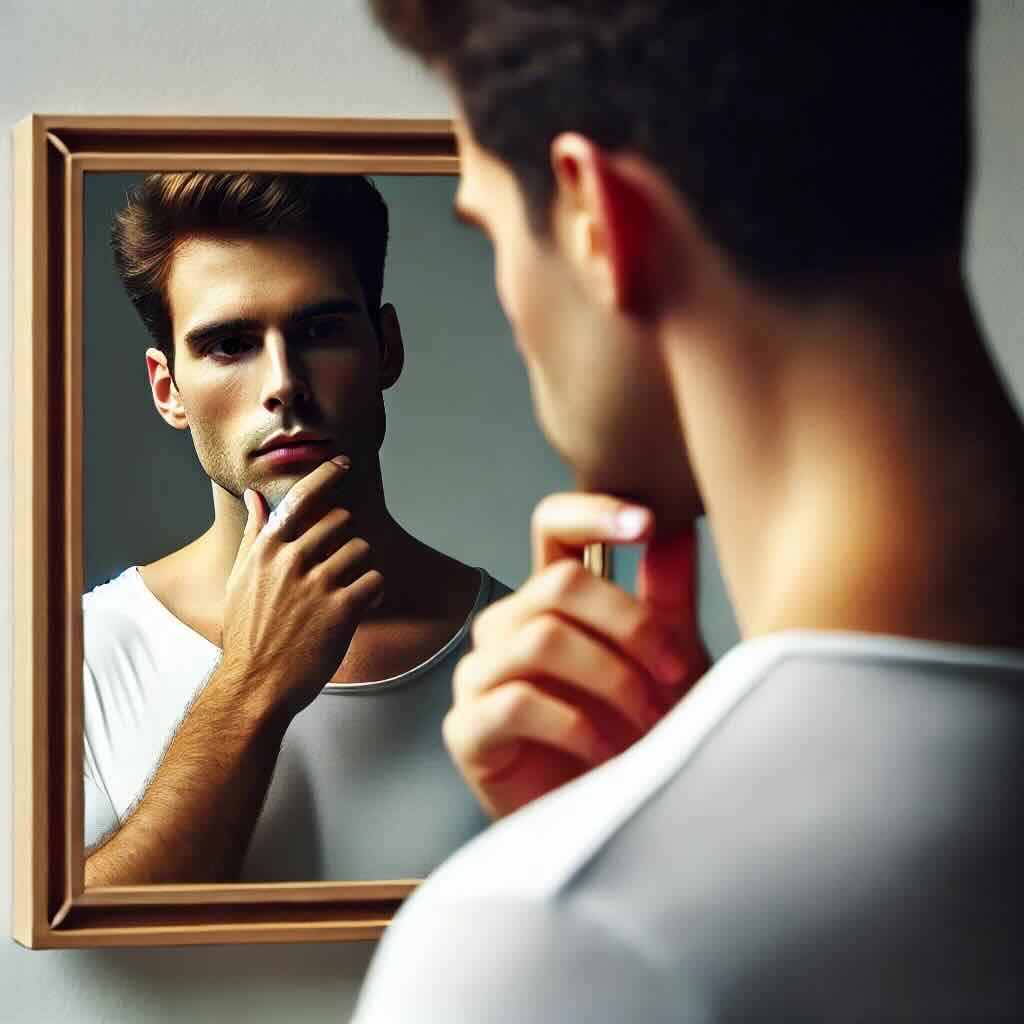}
& \includegraphics[width=0.15\textwidth]{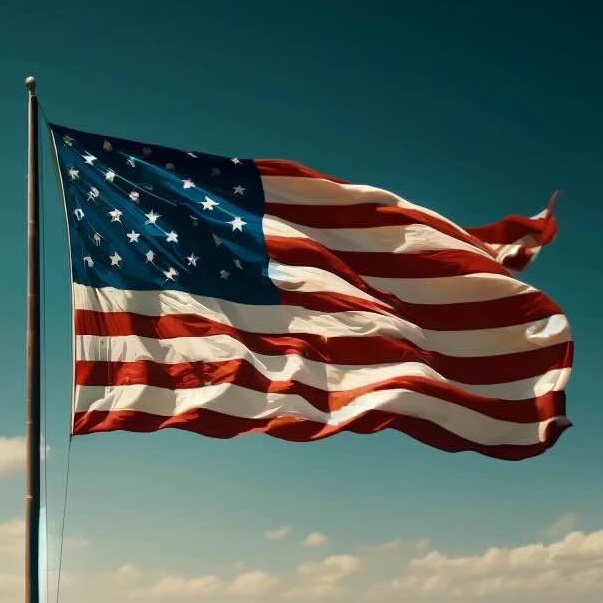}
& \includegraphics[width=0.15\textwidth]{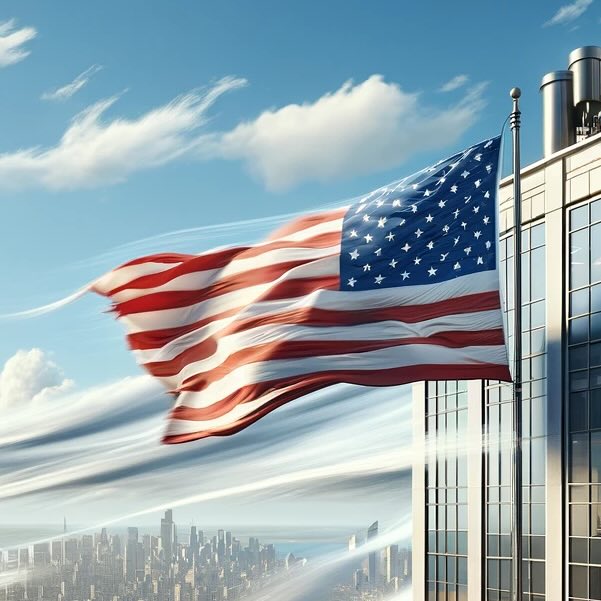}
& \includegraphics[width=0.15\textwidth]{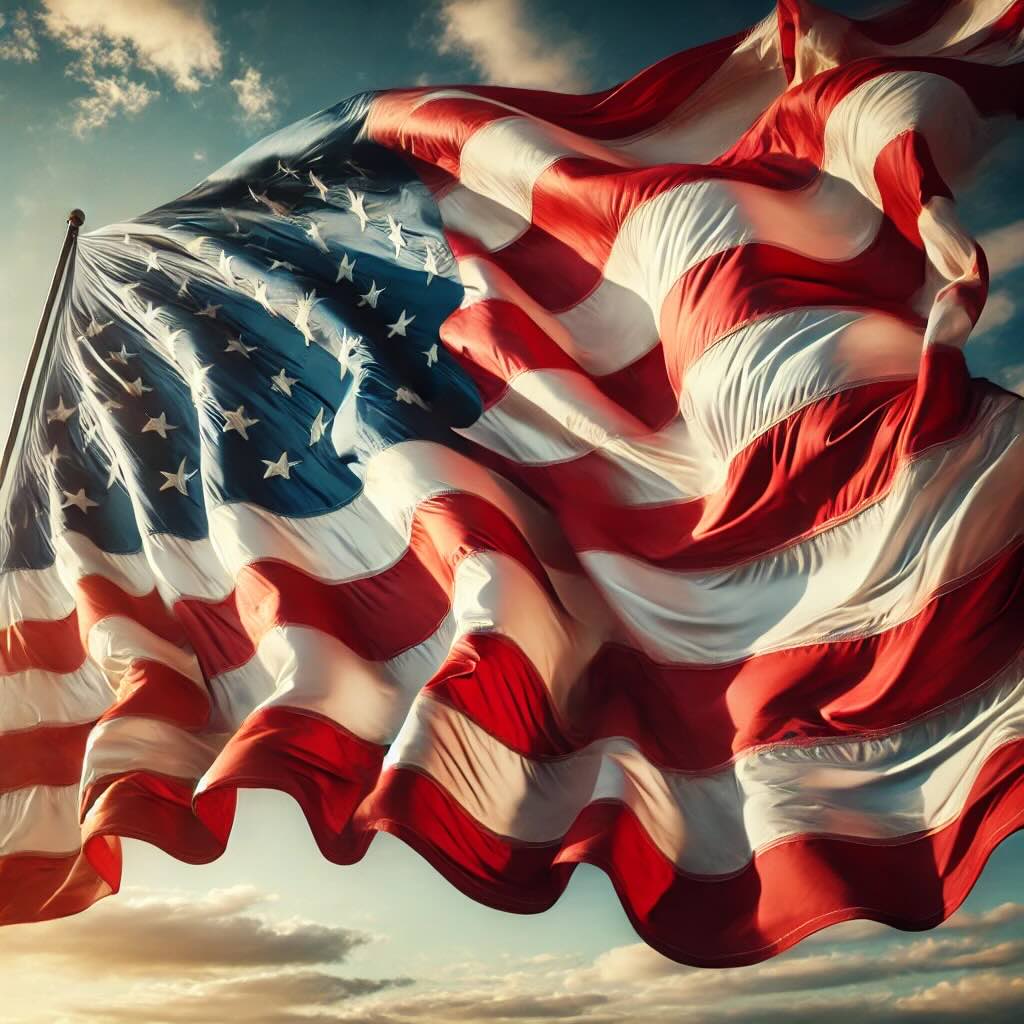}\\
 \hspace{.35cm} \rotatebox[origin=t]{90}{Midjourney}
& \includegraphics[width=0.15\textwidth]{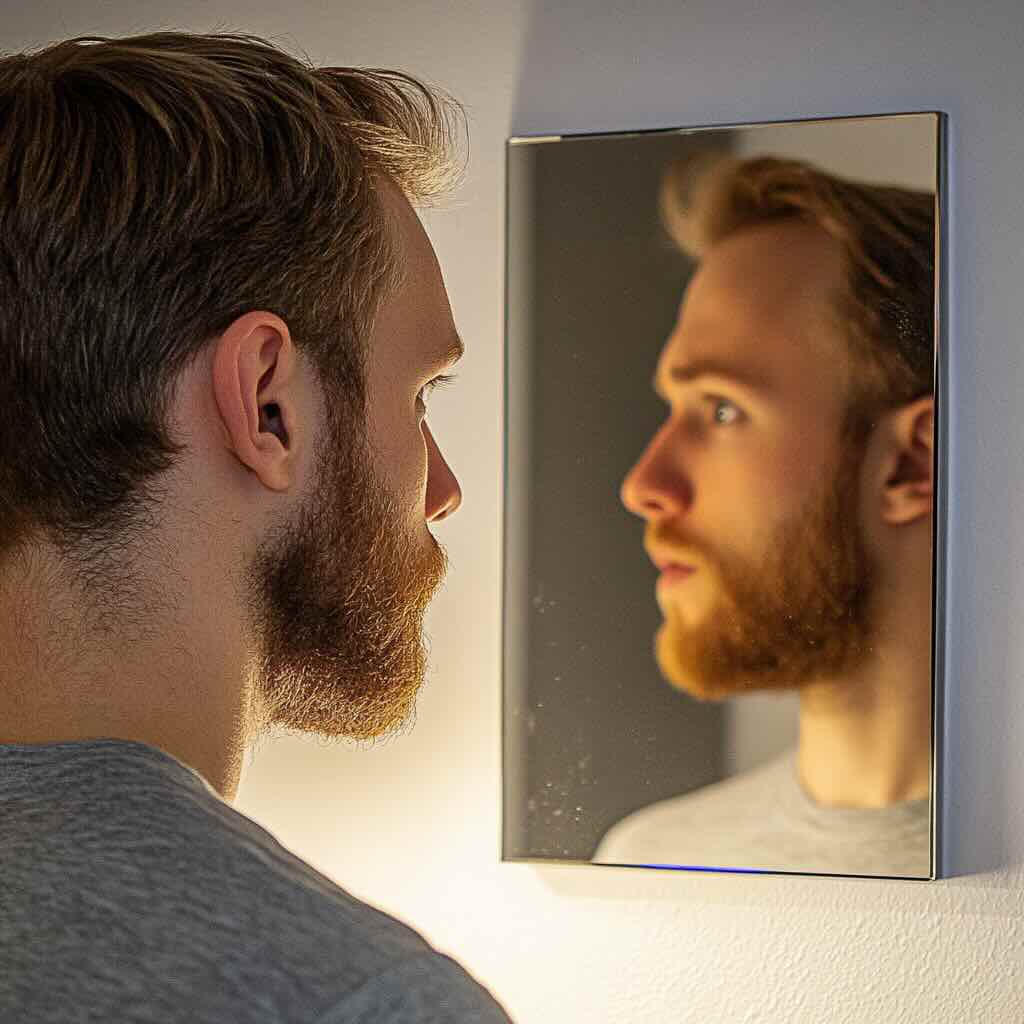}
& \includegraphics[width=0.15\textwidth]{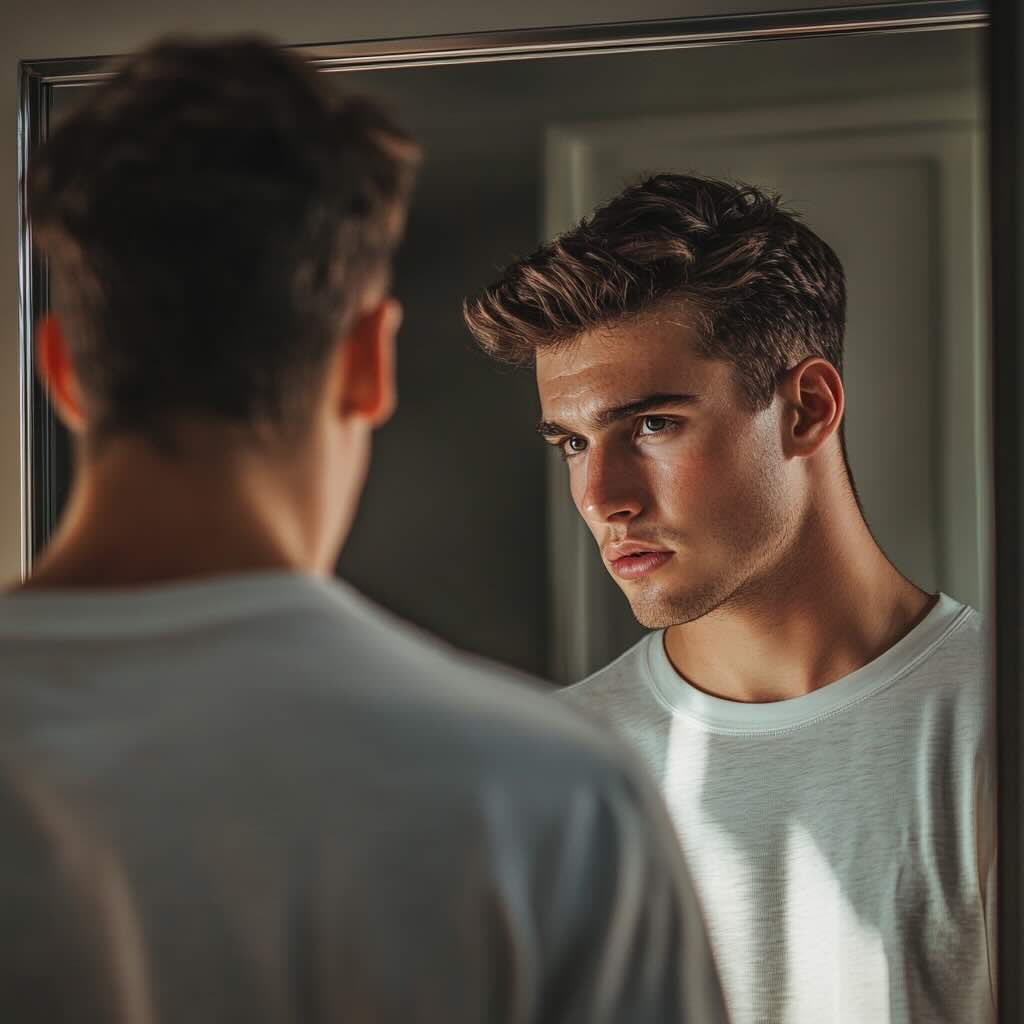}
& \includegraphics[width=0.15\textwidth]{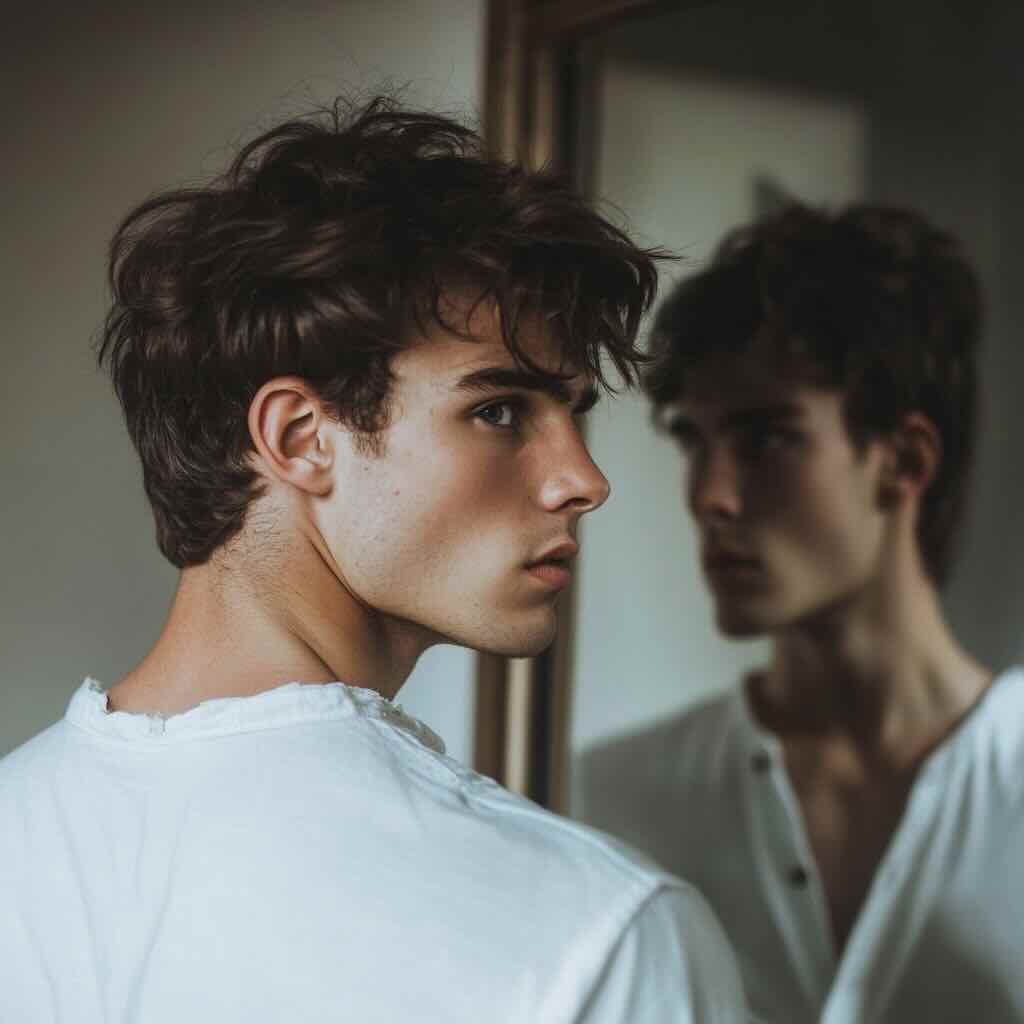}
& \includegraphics[width=0.15\textwidth]{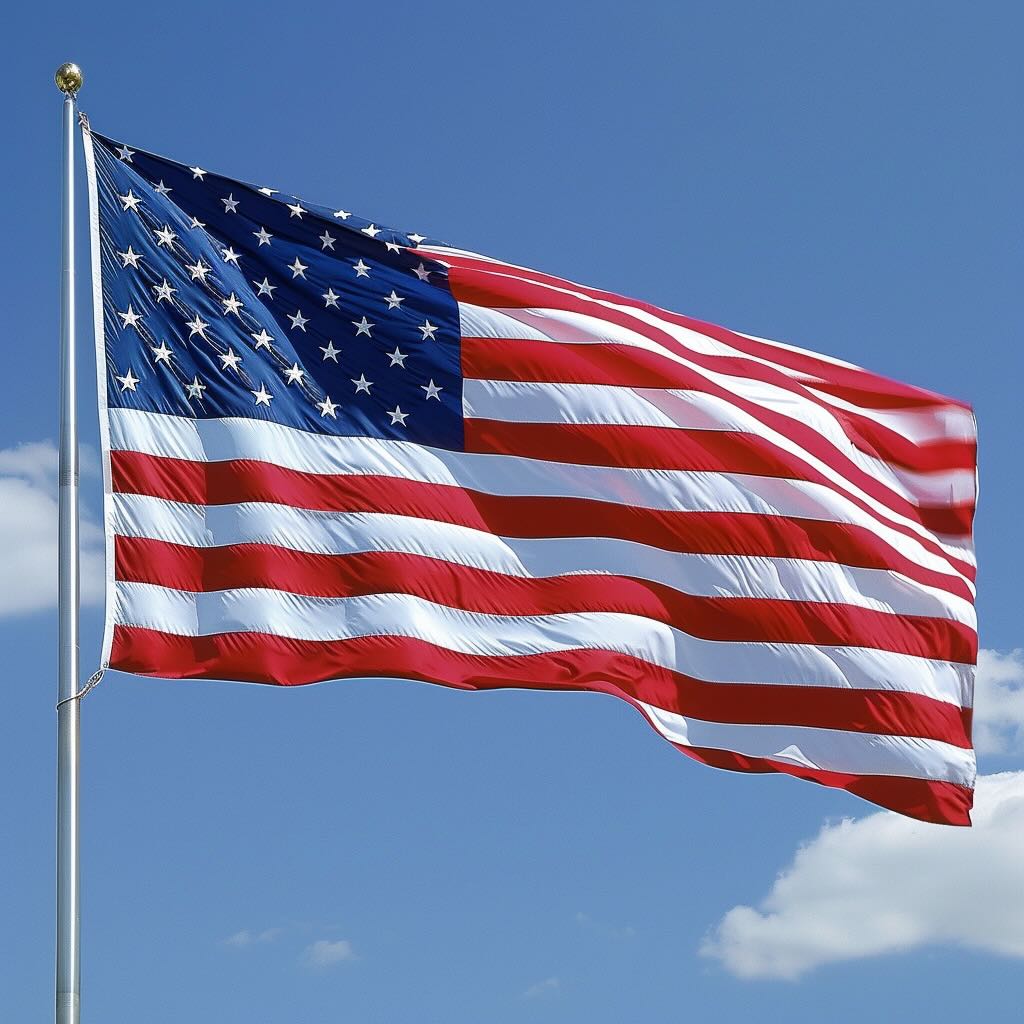}
& \includegraphics[width=0.15\textwidth]{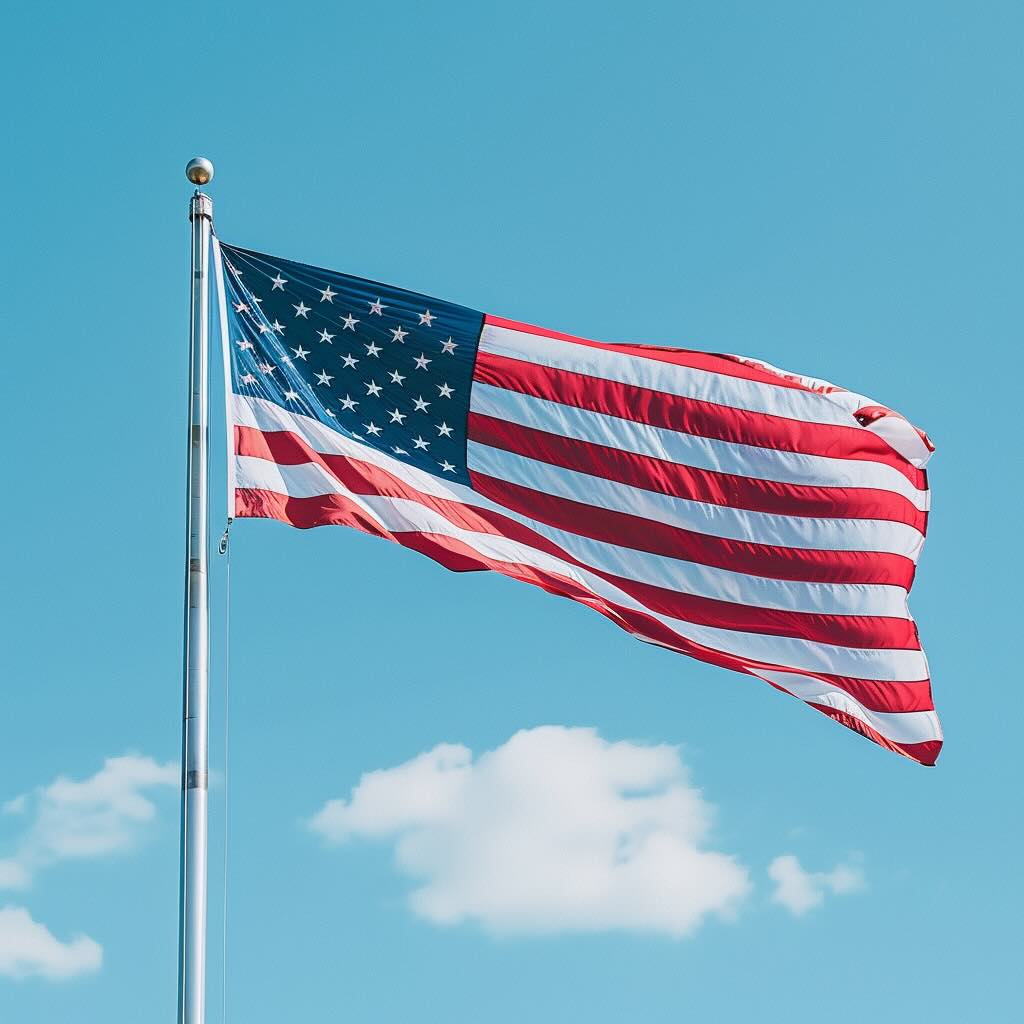}
& \includegraphics[width=0.15\textwidth]{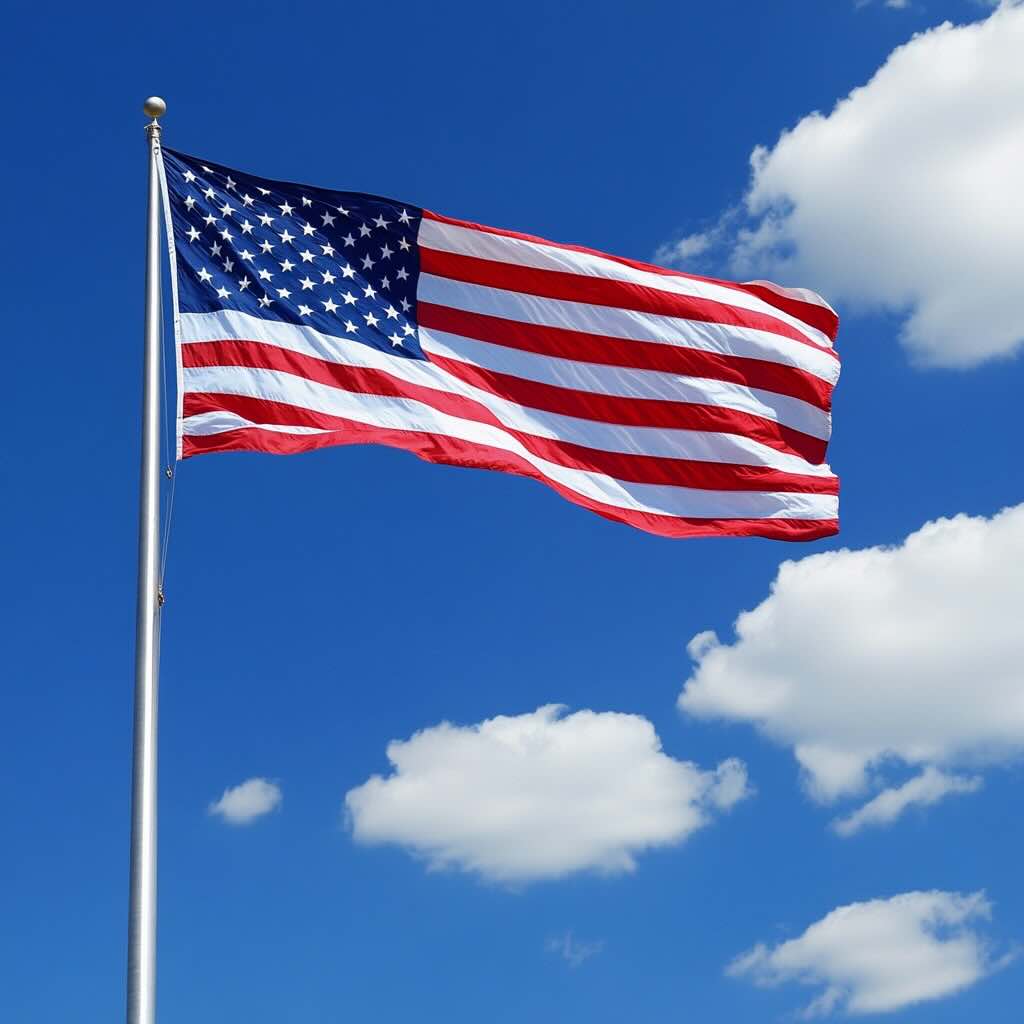}\\
 \hspace{.45cm}\rotatebox[origin=t]{90}{Meta AI}
& \includegraphics[width=0.15\textwidth]{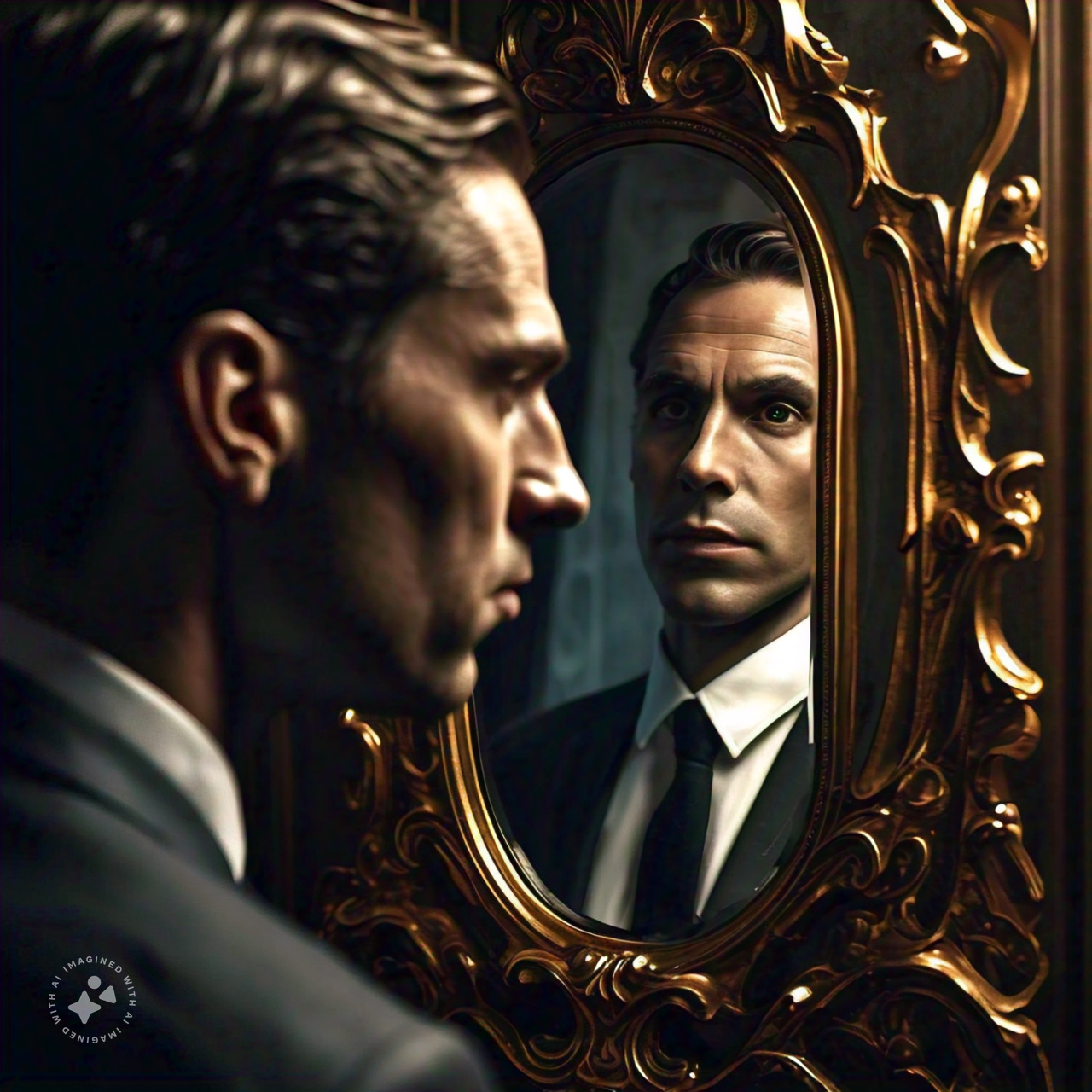}
& \includegraphics[width=0.15\textwidth]{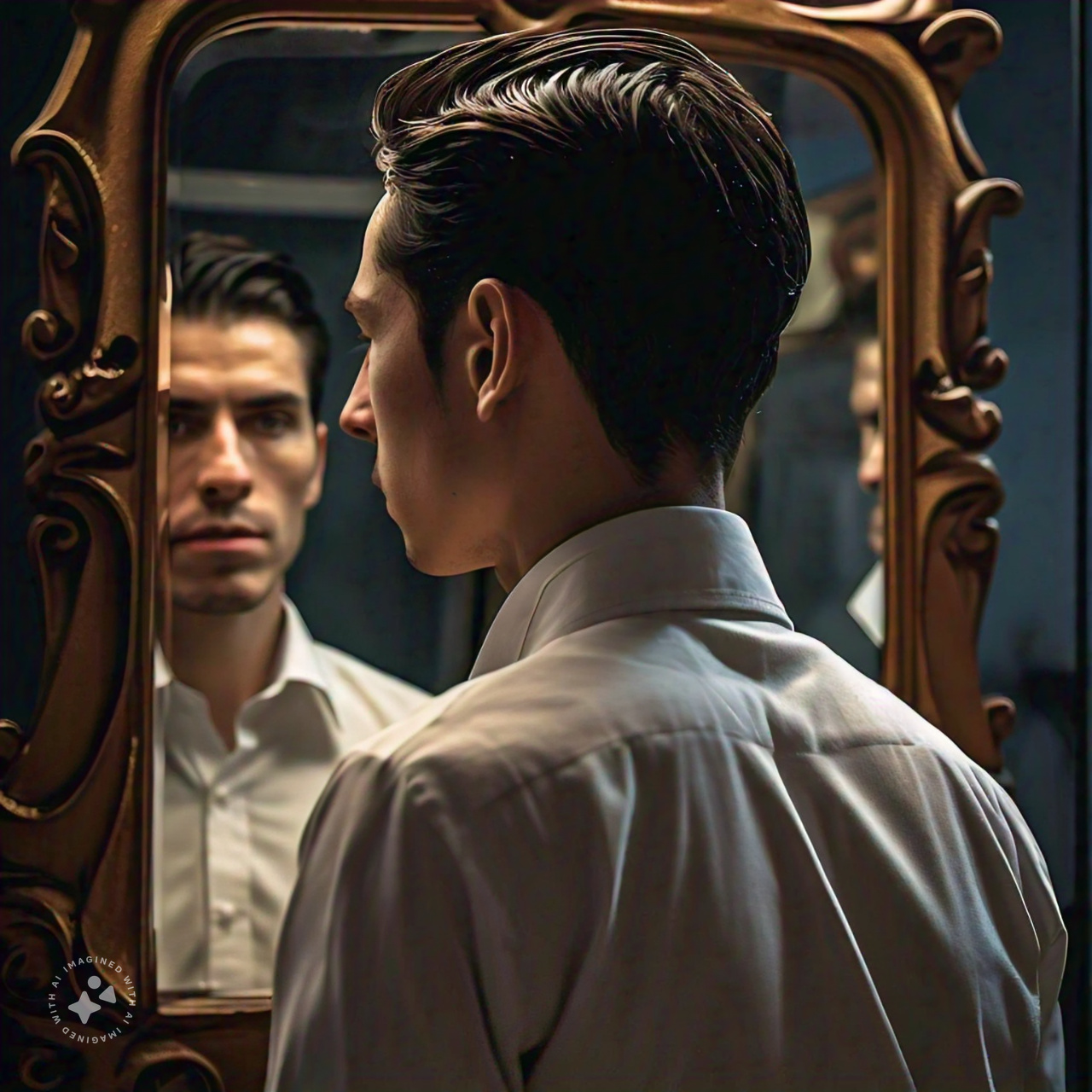}
& \includegraphics[width=0.15\textwidth]{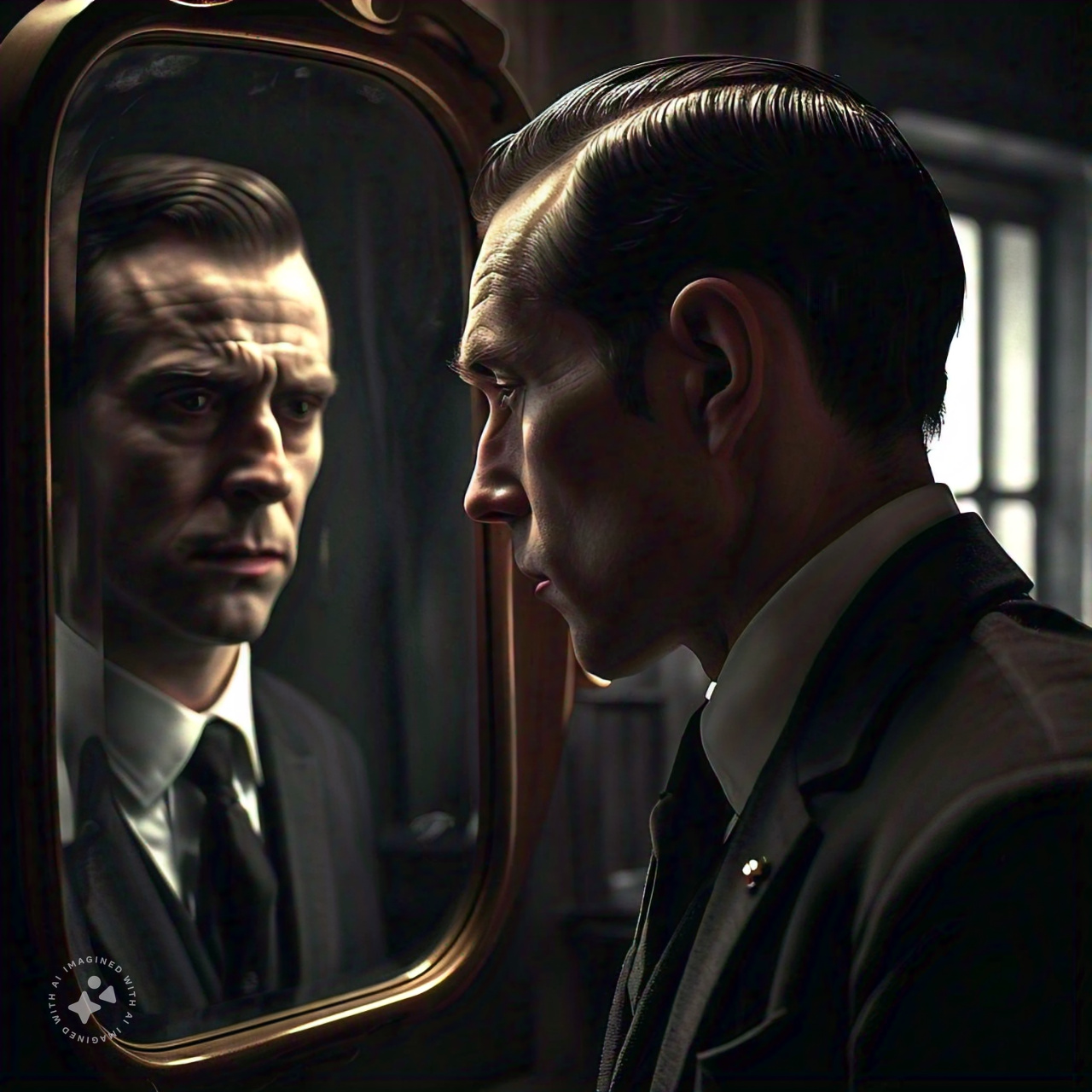}
& \includegraphics[width=0.15\textwidth]{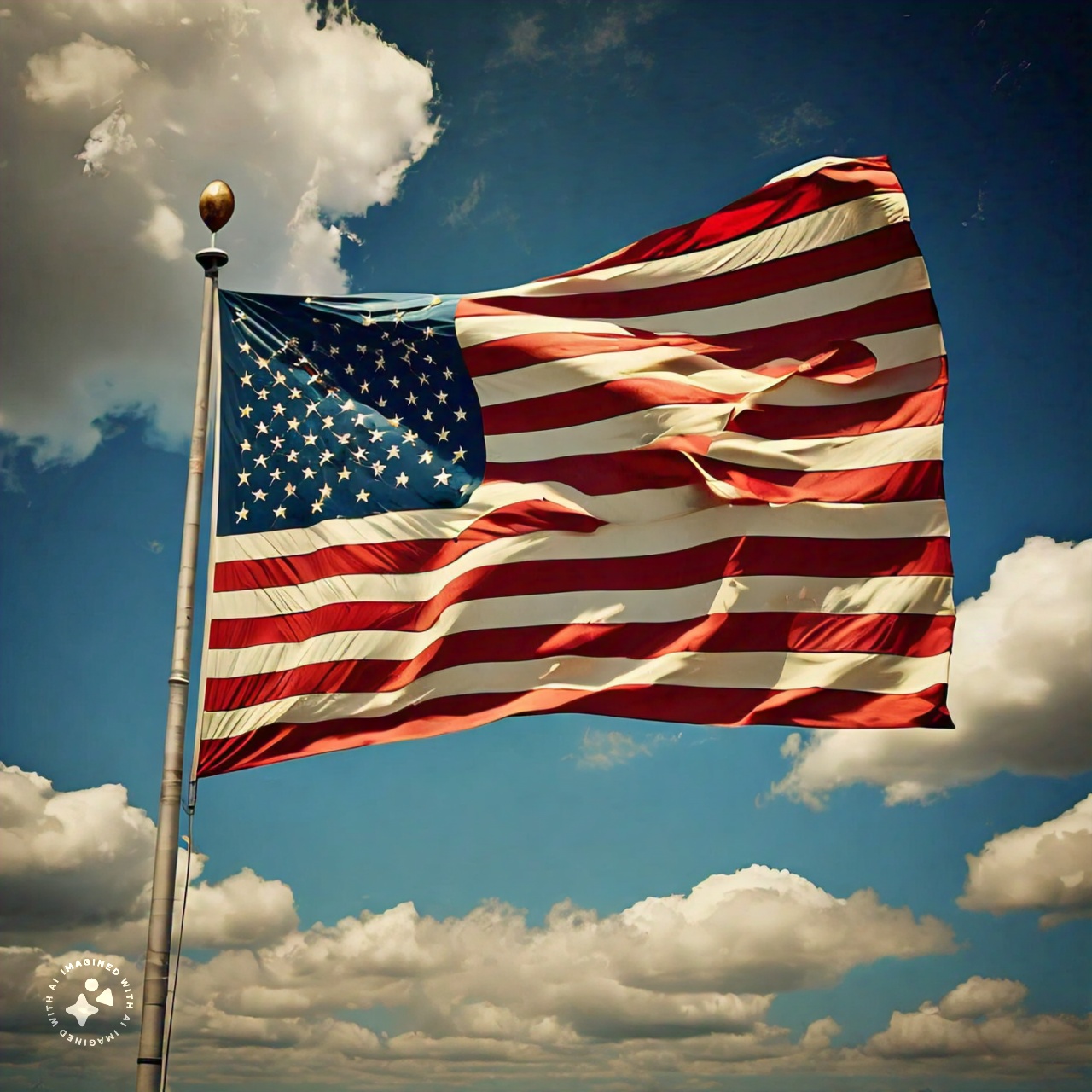}
& \includegraphics[width=0.15\textwidth]{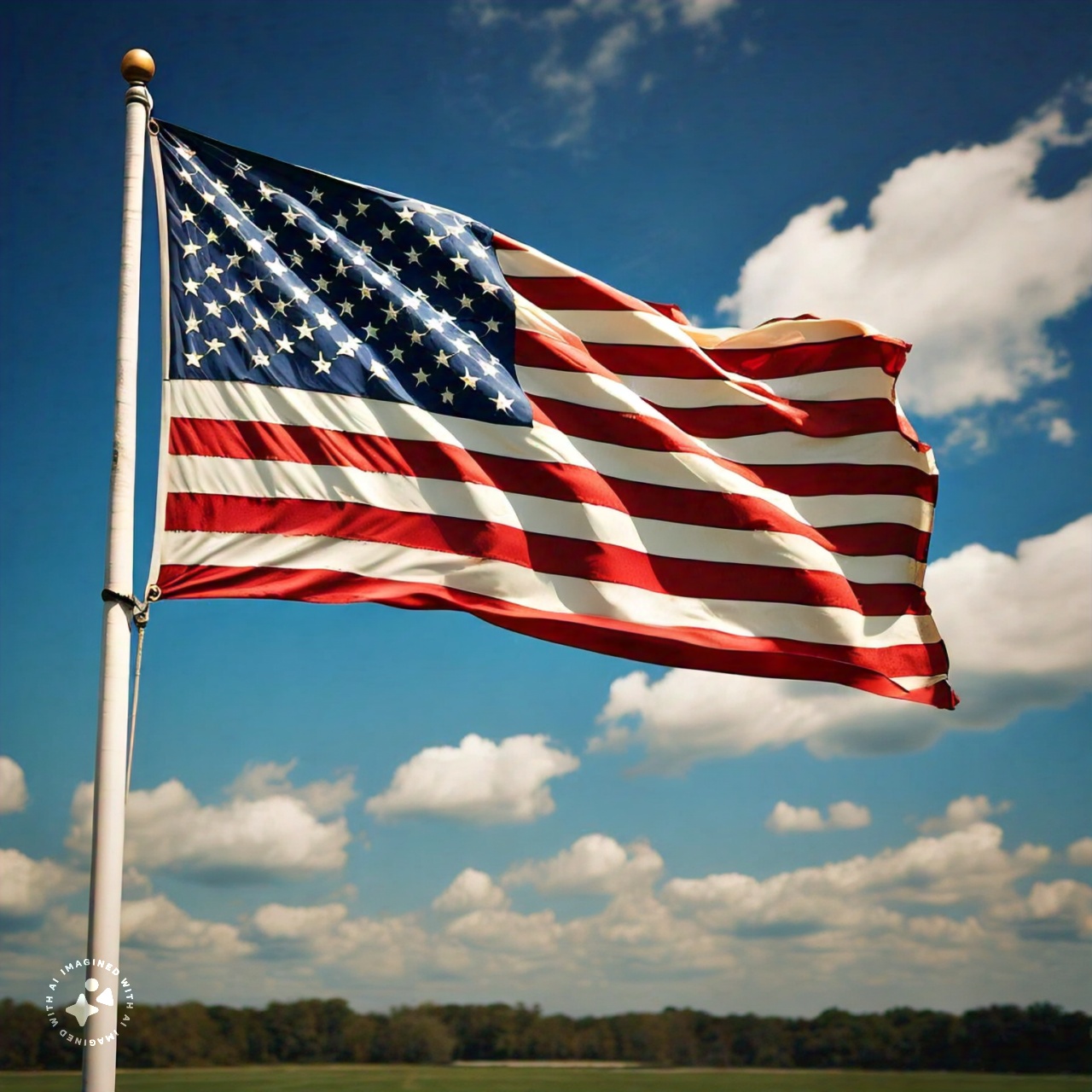}
& \includegraphics[width=0.15\textwidth]{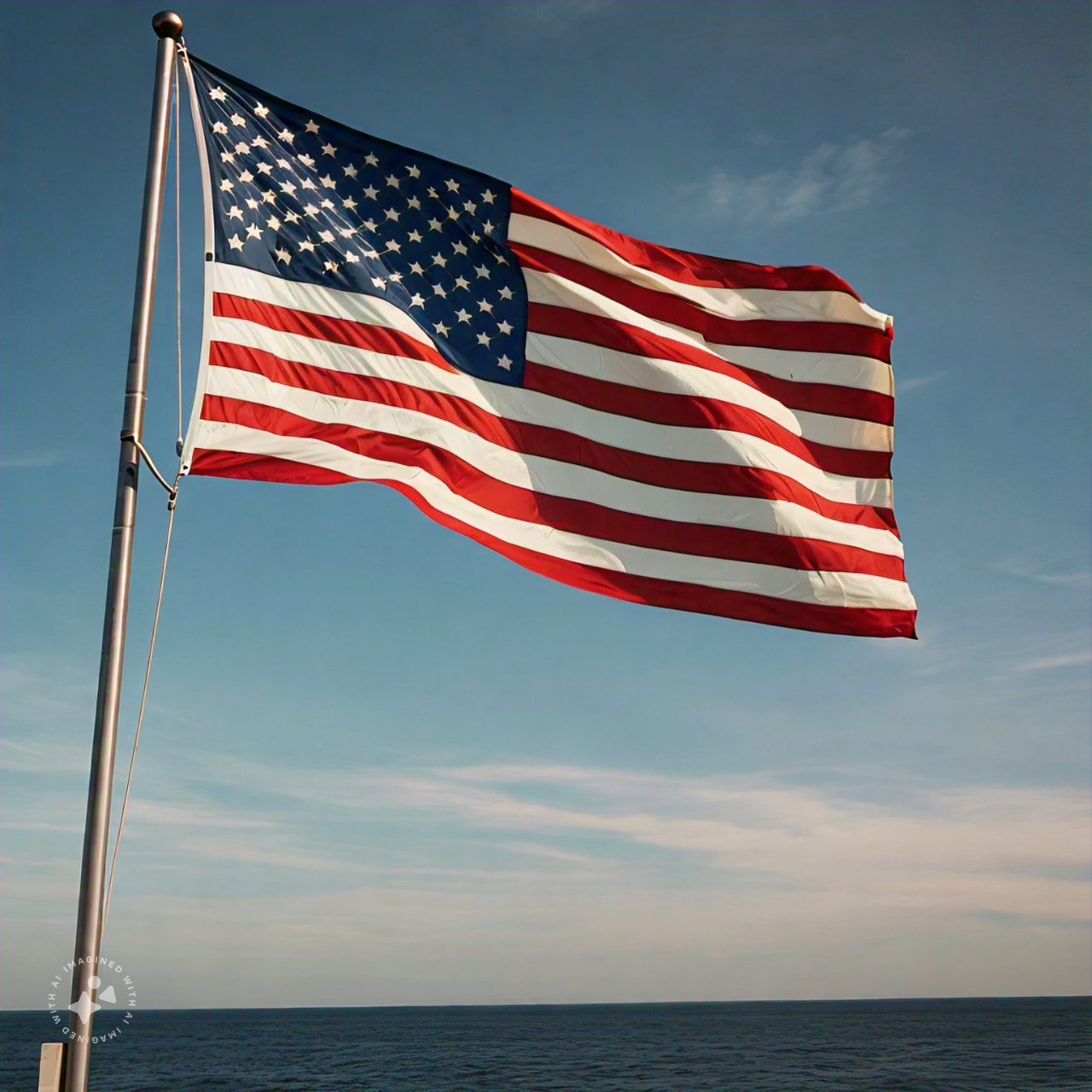}

\end{tabularx}
\caption{Illustrating the incoherence in images generated by Dall-E 3\protect\footnotemark, Midjourney\protect\footnotemark, and Meta AI\protect\footnotemark. These inconsistencies are evident in images featuring mirrors and wavering flags. Notice the forked or merged stripes on the flags and the inconsistent reflection and incidence angles in mirrors, among the other inconsistencies.}
\label{fig:vis_form}

\end{figure*}

Image generative models, such as those for drawing hands, struggle to generate coherent images due to the extensive variability in the training data \cite{narasimhaswamy2024handiffuser, zhang2024hand1000}. This issue extends to other structures with a high degree of variation. Figure \ref{fig:vis_form} illustrates two other structures (mirrors and wavering flags) with significant variation that three commercial image generators (Dall-E 3 \cite{dalle}, Midjourney, and Meta AI) fail to generate coherent images for. These generated images support the argument that learning coherent structure in the presence of large variability in the data manifold is a practical limitation. A solution that has been explored with promising results is reducing the variability by conditioning the model on a extra signal, for instance using predicted 3d hand pose from text prompt to generate images with hands \cite{narasimhaswamy2024handiffuser}. Our method follows a similar approach.

\newcolumntype{R}{@{\extracolsep{5cm}}r@{\extracolsep{0pt}}}%
\newcolumntype{L}{l<{\hspace{.15cm}}}
\newcolumntype{F}{l<{\hspace{.0cm}}}

\footnotetext{\url{https://chatgpt.com/}}
\footnotetext{\url{https://midjourney.com/}}
\footnotetext{\url{https://meta.ai/}}

To illustrate the problem that generative models struggle with learning musical form, consider a simple case. A generative model using maximum likelihood estimation optimizes $\theta$ to estimate the joint probability $p_{\theta}(t_1,t_2)$ from the data samples with discrete and finite values. A parametric model can estimate the joint probability if the training data samples lie on a compact manifold. If there is a large amount of variation in $t_1$ among the samples, then $p_{\theta}(t_1,t_2)$ reduces to $p_{\theta}(t_2)$ as $p(t_1)$ becomes virtually uniform. With high dimensional data, the problem of non-compact data manifold is compounded with the fact that with an arbitrarily good model as judged by average log-likelihood, $-log[p_{\theta}(t_1,\dots,t_d)] \approx d$. The optimization process minimizes the negative log-likelihood by adjusting $\theta$ for the dimensions with less variation. In a general musical corpus, with many parameters varying across music pieces with the same musical form, the amount of variation at large temporal scales to learn musical form is large enough that even simple musical forms are impossibly difficult to learn for generative models.

\section{MusicGen}

The most recent methods in the literature for music generation follow the approach of Stable Diffusion \cite{rombach2022high}. In this approach, the generative model is trained in the latent space of an encoder. The generated vector is then decoded into audio using the corresponding decoder. The main reason for training the generative model in the compressed latent is to reduce the computational cost of generating long audio. In MusicGen, an autoregressive model is trained in the quantized latent space of EnCodec~\cite{defossez2022high} to model music. 

To condition the MusicGen on text, the text prompt is encoded by T5 \cite{raffel2020exploring}. Classifier-free guidance (CFG) is used to generate samples with text conditioning.  MusicGen is trained on 20K hours of music. Half of the training data is private and internal at Meta. The other 10K hours of training music is taken from ShutterStock\footnote{\url{https://www.shutterstock.com/music}} and Pond5\footnote{\url{https://www.pond5.com}} vocals-free music data collections.

\section{Controlling MusicGen by a Language Model}

As MusicGen is conditioned on text, it affords an interface in natural language. Thus, an LLM can generate prompts for MusicGen to replace human prompts. Therefore, it is possible to task an LLM to design the structure of a song and generate prompts for each part to be generated by a text-to-music model. This capability of an LLM to design music structure and generate prompts is supported by its diverse knowledge base~\cite{schulze2024empirical}, the reasoning~\cite{webb2023emergent}, and learning capabilities~\cite{brown2020language}.

The first challenge is aligning the LLM with the text-to-music model. MusicGen has has been trained on brief music descriptions that are not technical and adhere to a certain style.  The LLM needs to generate prompts that MusicGen can interpret. To bridge this gap, there are two main approaches.One is fine-tuning or training a pre-trained LLM, which involves adapting it directly to the task. Another approach is in-context learning, which offers the advantages of requiring fewer samples and being less resource-intensive when trying and evaluating different prompts. Notably, using a large number of samples for in-context learning has been shown to outperform fine-tuning in terms of accuracy \cite{agarwal2024many,bertsch2024context}.

In-context learning is used to instruct ChatGPT to generate prompts for MusicGen by providing 50 song descriptions from Pond5. To find the required number of samples, it is estimated how often the generated music is faithful to the text prompt to MusicGen. The empirical results show that with about 10 song descriptions the generated by prompts were often misinterpreted by MusicGen. Increasing the number of samples from 50 to 80 showed no improvement in the interpretability of the prompts by MusicGen. As LLMs are more prone to hallucination~\cite{fang2024large}, more than necessary samples should be avoided. 

ChatGPT tends to mix multiple genres in the prompts for a single piece, resulting in structures that are not particularly well-designed. However, when ChatGPT is directed to consider a framework such as the ITPRA theory~\cite{huron2008sweet}, the resulting structures are more organized and coherent. Additionally, employing the chain of thought approach~\cite{wei2022chain}—asking it to first respond with a description of a song and its form, followed by generating prompts for the parts—further improves the organization of the musical structure. Additional rules are added in the prompt to ChatGPT to introduce constraints specific to MusicGen. For instance, since MusicGen cannot generate parts that have "a slower tempo" than a previous part, ChatGPT is instructed to generate the prompts, the length of each part in seconds, and the musical elements from prior parts that should be used, all formatted in JSON. In the following prompt excerpt, some sections have been omitted for brevity.

\begin{lstlisting}[frame=single]
*Task* Assume you're a musician. Your task is to write text prompts for a system that generates music based on the given description of the music...
*Multishot examples* Below are some example prompts that the system understands, along with the type of music it can generate: 
- "a light and cheerful EDM track with syncopated drums, airy pads, and strong emotions; bpm: 130" 
...
*Constraints* Don't limit yourself to these example prompts... The music piece should be coherent and have a sense of unity. Describe your thought process for the composition, followed by the breakdown of the different parts... The following are important constraints that your prompts must satisfy: 
1. The entire piece must be exactly 150 seconds long. You will also decide the length of each part. 2. The prompt for each part can reference another part... 14. To repeat a part with variations in the chosen musical form, reference the original part and, in the new prompt, explain what changed.
*Request Part 1* Come up with 10 pieces, including the form and a description of each part.
*Request Part 2* ... Provide the details of the parts for each piece in JSON format: { PART_NUMBER: ["PROMPT", LENGTH_IN_SECONDS, REFERENCED_PART], PART_NUMBER: ["PROMPT", LENGTH_IN_SECONDS, REFERENCED_PART], ... }
\end{lstlisting}

\section{Evaluation}
\begin{figure}[!t]
\begin{center}
\includegraphics[width=0.99\columnwidth]{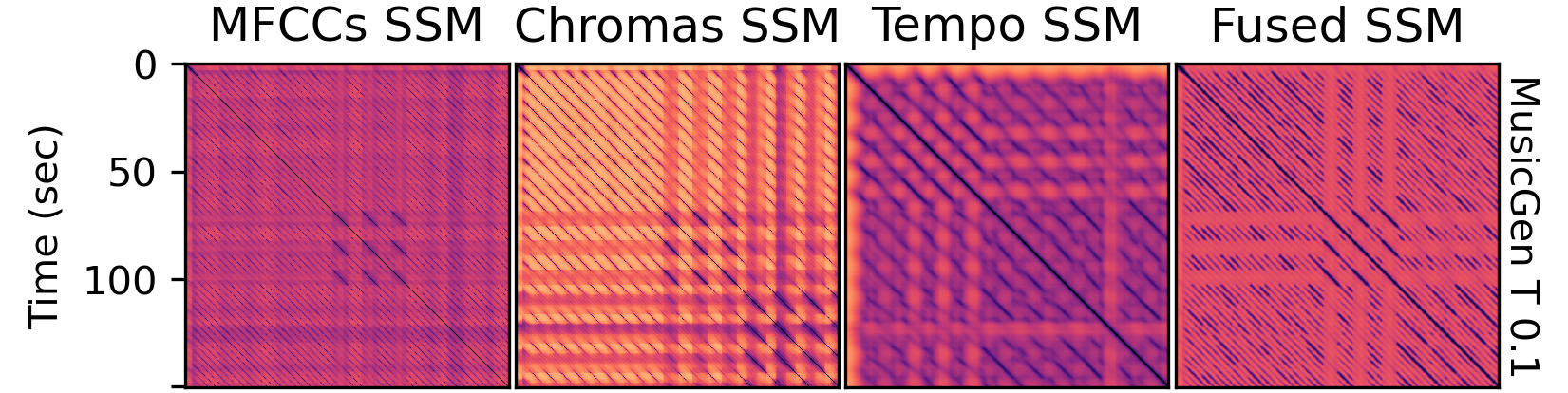}
\includegraphics[width=0.99\columnwidth]{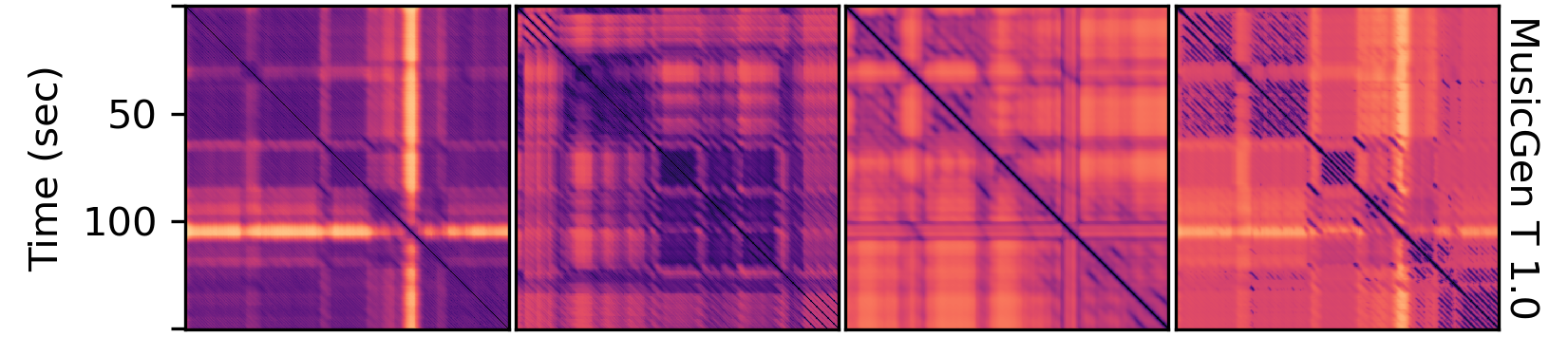}
\includegraphics[width=0.99\columnwidth]{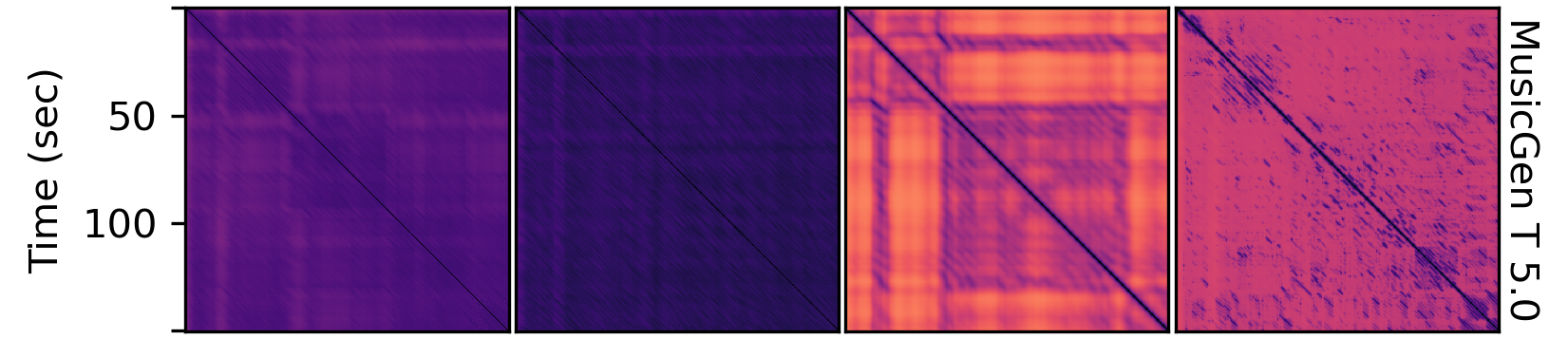}
\includegraphics[width=0.99\columnwidth]{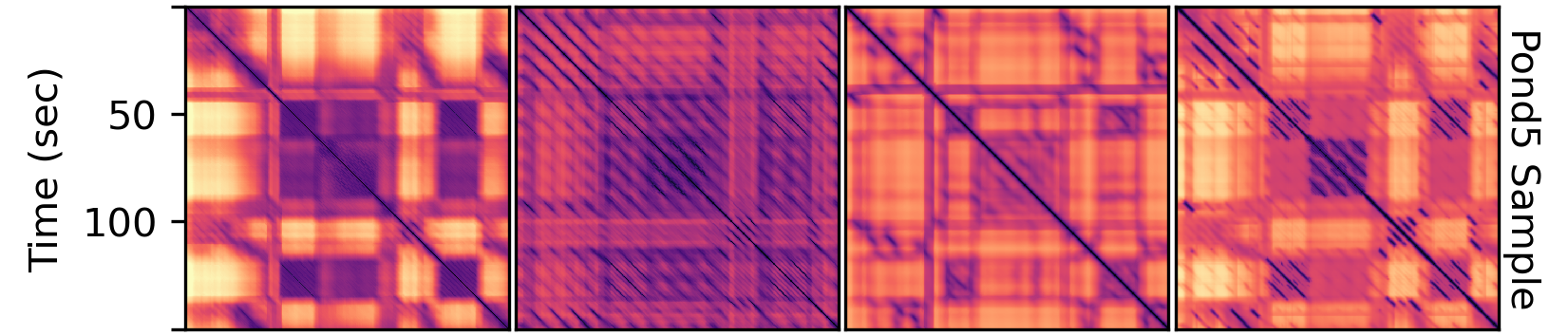}
\includegraphics[width=0.99\columnwidth]{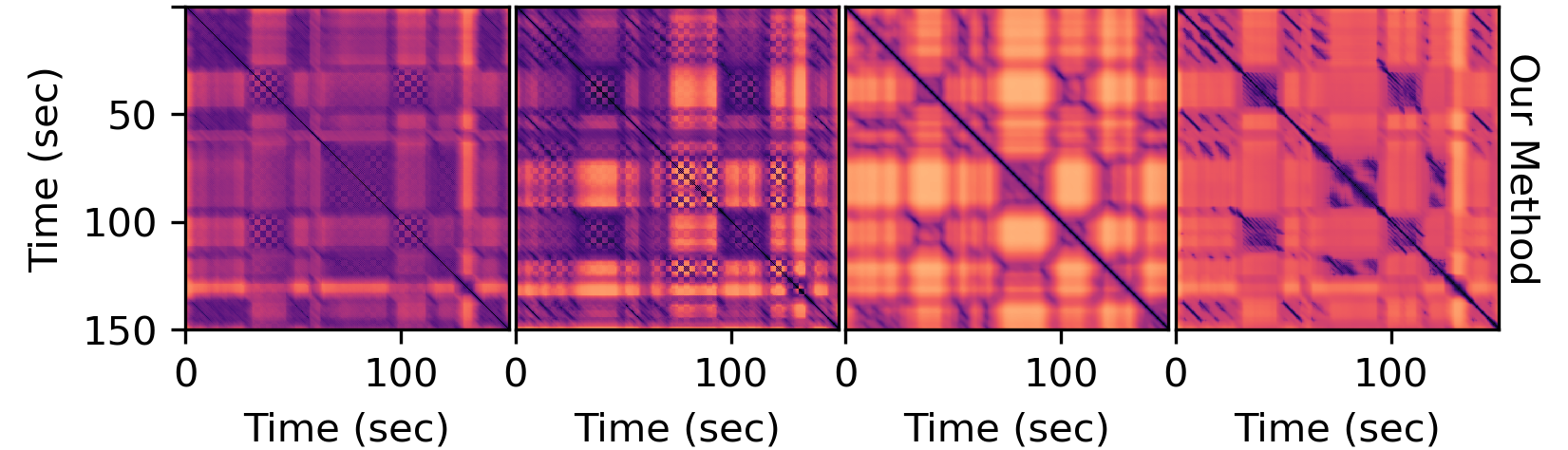}

\caption{Visualizing the self similarity matrices for 3 MusicGen samples, one sample from our method and one from Pond5. With MusicGen, at a low temperature (T) of 0.1, the music is repetetive. At T=5.0, there is mostly random noise. At T=1, the music is meandering. The sample from our method resembles the one from Pond5, composed and arranged by a musician.}
\label{fig:musicgen_structure}
\end{center}
\end{figure}

Switching between prompts passed to MusicGen creates a sudden jump. To ensure a smooth transition between parts, the CFG method is modified. Instead of estimating one conditioned probability distribution, two conditioned distributions are estimated. In the interpolated probability distribution, the weight of the first conditioned distribution over five seconds goes from one to zero linearly, and the weight of the second conditioned distribution increases from zero to one. This method allows the model to generate tokens that facilitate a continuous transition between parts.

Similarly, to generate parts with variations of the prior parts, a 15-second audio prompt from a previous part is provided to MusicGen. This 15-second audio is passed to the encoder of EnCodec to generate the corresponding tokens. Then, the prompt tokens are passed to the predictive model of MusicGen. The weight of the probability distribution over tokens, conditioned on these audio prompt tokens, is then linearly reduced to zero over 10 seconds.

Figure~\ref{fig:musicgen_structure} compares the self similarity matrices (SSM) of samples from MusicGen, our method, and Pond5. The fusion method proposed in ~\cite{tralie2019enhanced} is used to generate the combined SSMs. The SSMs of the sample from our method resemble the variation and similarity structure in the sample from Pond5. 

After downsampling the fused SS matrices of 100 samples, the mean and variance matrices are estimated for Pond5, our method, and MusicGen in Figure~\ref{fig:mus_objective_eval}. Therefore, this figure helps to compare the distributions of the music structures. It is apparent that, from the mean and variation values from the top-right of the matrices, the MusicGen samples do not have parts near the end of each piece that resemble the parts near the beginning. In contrast, the samples from our method are more similar to the samples from Pond5. The Fr\'echet distance between the two pairs of distributions is estimated, but instead of using Inception feature vectors \cite{heusel2017gans}, the mean and covariance matrices of the upper triangle of the fused SS matrices are used in the distance estimation. The Fr\'echet distances between the the distributions of the Pond5 samples and ours is 0.086, and between the Pond5 and MusicGen samples is 0.108, supporting the claim the structure of the samples generated by our method are more similar to the samples composed by musicians from Pond5. 

\begin{figure}[!t]
\begin{center}
\includegraphics[width=0.49\columnwidth]{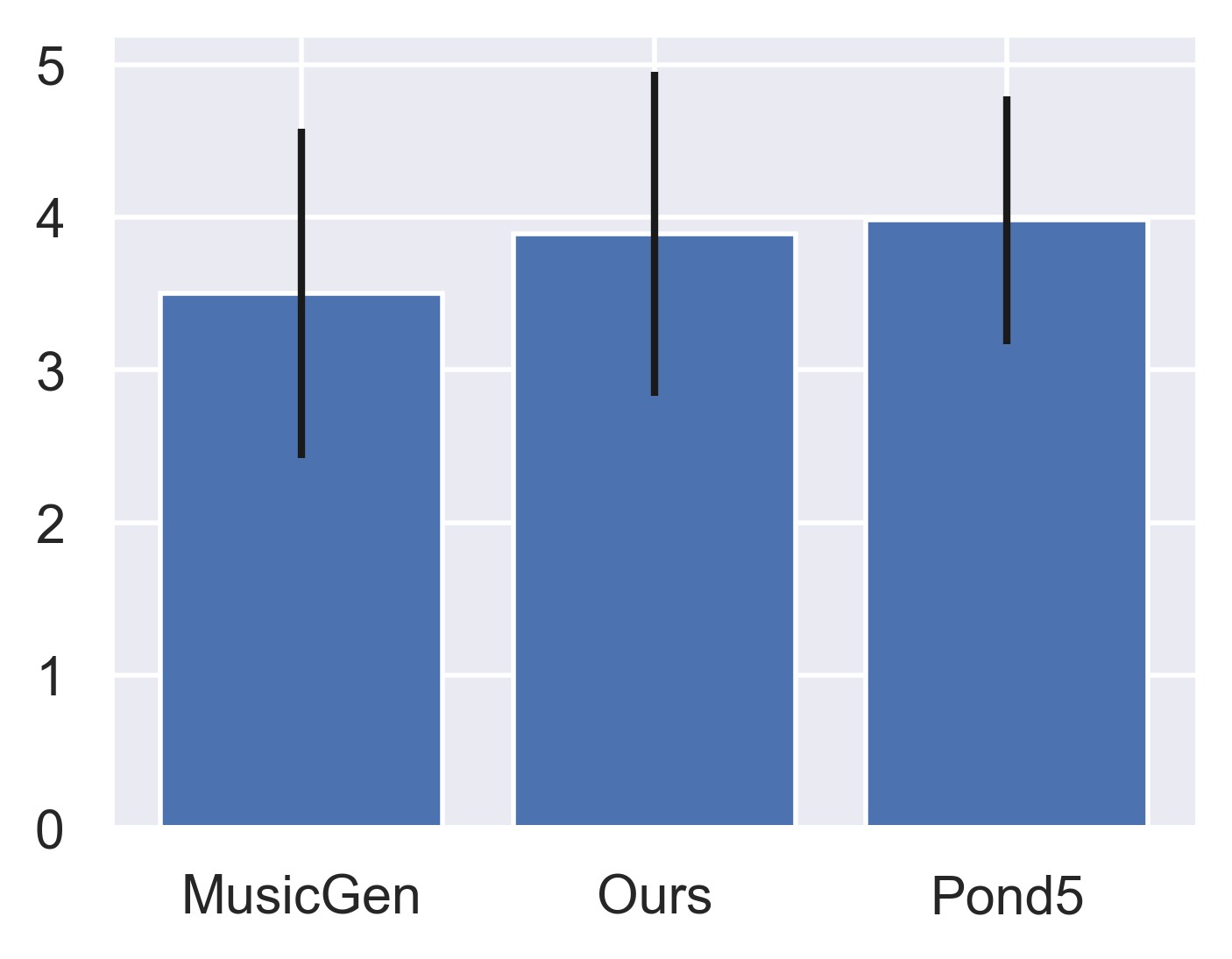}
\includegraphics[width=0.49\columnwidth]{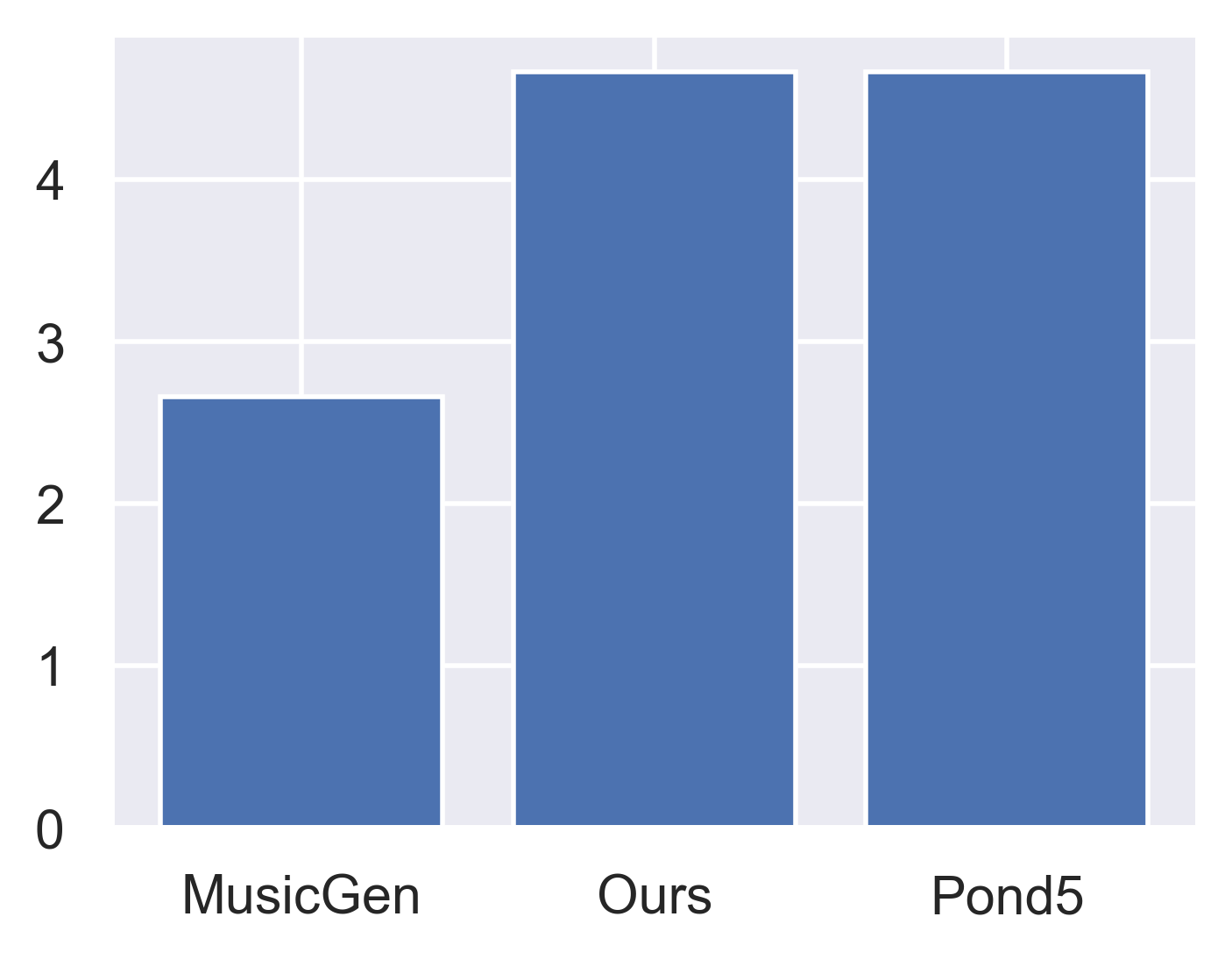}

\caption{Left: The subjective comparison of the generated music and sampled from Pond5 by non-musicians is measured through the MOS based on how engaging the music is. Whiskers: 95\% CI. Right: The subjective comparison of the samples by musicians, critiquing the musical structures.}
\label{fig:subjective_eval}
\end{center}
\end{figure}

For a subjective evaluation by non-musicians, 10 samples are generated using our method, 10 samples using MusicGen, and 10 samples from Pond5. Each of these samples is 2.5 minutes long. Using Amazon Mechanical Turk (MTurk), a mean opinion score (MOS) between 1 and 5 for each sample from 10 non-musicians is collected. The human raters are asked to evaluate the overall quality of the music, following the recommended practices in CrowdMOS~\cite{ribeiro2011crowdmos}. The subjects are told a score of 1 means they dislike the music and would not like to listen to similar music. A score of 5 means they find the music interesting and would like to listen to similar music. Given long music tracks with more organized structure are expected to be more engaging, the likeness of the samples is used as a proxy for the improved structure and form. The results in Figure~\ref{fig:subjective_eval}(top) indicate that adding musical form through our method to MusicGen improves the perceived quality of the music, almost on par with human-composed music pieces from Pond5.

In a separate subjective evaluation, three professional musicians with doctoral degree in performance or composition are asked to listen to three samples from our method and three samples from MusicGen. 
They assign a score to each sample, with the score guideline as follows: 1: No form, music meanders; 2: Minimal structure, some recognizable patterns but largely unstructured; 3: Moderate structure, clear sections but not highly organized; 4: Clear form, well-organized sections and transitions; 5: Very clear and highly structured, strongly organized and cohesive. The average scores are presented in Figure~\ref{fig:subjective_eval}(bottom).

\section{Conclusions}

This paper argues that due to the discussed nature of the problem, the music generative models cannot learn long-scale structure or musical form from a musical datasets. Then a novel method to combine music generative models with LLMs to generate music with a musical form is presented. The technical challenges are discussed. The subjective and objective evaluations support the claim that our method can generate well-structured 2.5-minute music.

\bibliographystyle{IEEEbib}
\bibliography{main}

\end{document}